\documentclass[preprint2]{aastex}
\usepackage{textcomp}
\usepackage{amsmath}
\usepackage{float}
\slugcomment{To appear in ApJ.}

\shorttitle{Solar Wind Model at New Horizons}
\shortauthors{Kim et al.}

\begin{document}


%\linenumbers

\title{Modeling the Solar Wind at the Ulysses, Voyager, and New Horizons Spacecraft}


\author{T. K. Kim\altaffilmark{1}, N. V. Pogorelov\altaffilmark{1,2}, G. P. Zank\altaffilmark{1,2}, H. A. Elliott\altaffilmark{3}, and D. J. McComas\altaffilmark{3,4}}
%\affil{Center for Space Plasma and Aeronomic Research, University of Alabama in Huntsville, Huntsville, AL 35805, USA}

%\and

%\author{H. A. Elliott, D. J. McComas\altaffilmark{2,3}}
%\affil{Southwest Research Institute, San Antonio, TX 78249, USA}


\altaffiltext{1}{Center for Space Plasma and Aeronomic Research, University of Alabama in Huntsville, Huntsville, AL 35805, USA}
\altaffiltext{2}{Department of Space Science, University of Alabama in Huntsville, Huntsville, AL 35805, USA}
\altaffiltext{3}{Southwest Research Institute, San Antonio, TX 78238, USA}
\altaffiltext{4}{Princeton Plasma Physics Laboratory, Princeton University, Princeton, NJ 08544, USA}

\begin{abstract}
%\begin{linenumbers}
The outer heliosphere is a dynamic region shaped largely by the interaction between the solar wind and the interstellar medium. While interplanetary magnetic field and plasma observations by the Voyager spacecraft have significantly improved our understanding of this vast region, modeling the outer heliosphere still remains a challenge. We simulate the three-dimensional, time-dependent solar wind flow from 1 to 80 astronomical units (AU), where the solar wind is assumed to be supersonic, using a two-fluid model in which protons and interstellar neutral hydrogen atoms are treated as separate fluids. We use 1-day averages of the solar wind parameters from the OMNI data set as inner boundary conditions to reproduce time-dependent effects in a simplified manner which involves interpolation in both space and time. Our model generally agrees with Ulysses data in the inner heliosphere and Voyager data in the outer heliosphere. Ultimately, we present the model solar wind parameters extracted along the trajectory of New Horizons spacecraft. We compare our results with in situ plasma data taken between 11 and 33 AU and at the closest approach to Pluto on July 14, 2015.
%\end{linenumbers}
\end{abstract}


\section{Introduction}

The New Horizons (NH) spacecraft joined Voyager and Pioneer to become the only interplanetary missions to explore the outer heliosphere. Launched on 19 January 2006, NH made its closest approach to Pluto at 32.9 AU on 14 July 2015 after more than 9 years of interplanetary travel. NH is equipped with a plasma instrument called Solar Wind Around Pluto (SWAP), which was designed to measure the solar wind and its interaction with Pluto's atmosphere \citep{McComas2008SSRv,McComas2016JGR}. The solar wind flow speed, density, and temperature parameters sampled by SWAP on its way to Pluto and beyond are a valuable addition to the limited database from this frontier region.

Pluto's orbit lies in the outer heliosphere where charge exchange between the solar wind and interstellar neutral atoms significantly affect the physical properties of the solar wind [see \cite{Zank2015ARAandA} and references therein]. For example, non-thermal ions called pickup ions (PUIs) generated in the charge exchange process decelerate the thermal solar wind flow with a drag provided by mass loading, and PUI-driven turbulence is believed to be responsible for the steady rise in solar wind temperature beyond $\sim$10 AU \citep{Matthaeus1999PRL,Isenberg2003ApJ,Isenberg2005ApJ,Smith2006ApJ}. Therefore, any model of the solar wind at distances greater than 10 AU should account for the charge exchange process and the creation of PUIs to reproduce the solar wind parameters reasonably.

The main objective of this work is to model the solar wind along the entire NH trajectory, which was inspired by the community-wide modeling effort to predict the solar wind parameters at NH for the Pluto flyby. We have compiled a set of time-dependent boundary conditions from near-Earth solar wind observations to drive our model that includes PUI effects. In the following sections, we describe the solar wind model and compare the simulation results with Ulysses and Voyager data to validate our model at various heliocentric distances and latitudes. Finally, we compare the model with NH-SWAP data between 11 and 33 AU and discuss how the model can be improved.

\section{Model}

The simulation utilizes the Multi-Scale Fluid-Kinetic Simulation Suite (MS-FLUKSS) that incorporates a hierarchy of physical models relevant to the interaction between the solar wind and local interstellar medium (LISM) \citep{Borovikov2013ASP,Pogorelov2014XSEDE}. The flow of plasma is described by solving the ideal magnetohydrodynamics (MHD) equations while neutral atoms can be treated either as multiple fluids by solving Euler gas dynamic equations \citep{Zank1996JGR,Pogorelov2009ASP} or kinetically by the Boltzmann equation \citep{Zank1996AIP,Heerikhuisen2008ASP}. MS-FLUKSS also includes PUI transport and supersonic solar wind turbulence models that allow separate treatment of PUIs, either kinetically or as a fluid, and heating of the solar wind in the outer heliosphere \citep{Pogorelov2011ASP}.

Originally designed to model the interaction between plasma and neutral atoms in the distant parts of the heliosphere and LISM, MS-FLUKSS has been used to investigate plasma instabilities at the heliopause \citep{BorovikovPogorelov2014ApJL} and the complex structure of the heliotail \citep{Pogorelov2015ApJL}, and also to provide the background plasma distribution for modeling the heliospheric modulation of galactic cosmic rays \citep{Zhang2014ApJ} and for reproducing the ``ribbon'' of energetic neutral atoms observed by the Interstellar Boundary Explorer (IBEX) spacecraft \citep{Zirnstein2016ApJL}. With time-dependent boundary conditions derived from Ulysses data, \cite{Pogorelov2013ApJ} analyzed the solar-cycle (SC) influence on the three-dimensional (3-D) structure of the outer heliosphere. As part of the recent efforts to implement more realistic boundary conditions based on observations to drive MS-FLUKSS, \cite{Kim2014ASP,Kim2014JGR} and \cite{Manoharan2015JPhCS} have experimented with boundary data from ground-based remote-sensing solar wind observations to reconstruct the inner and outer heliosphere. However, it will take time for these methods to mature. In this paper, we rely on OMNI data to provide time-varying boundary conditions. Dating back to the 1960s, OMNI data are updated regularly and are readily available at high cadence.

For this study, we employ a two-fluid model in which the plasma (both solar wind and PUI) are treated as a single magnetized fluid governed by MHD equations, whereas the flow of interstellar neutral hydrogen atoms is described by means of multi-component Euler equations. We use 1024 $\times$ 128 $\times$ 128 cells in a spherical coordinate system ($r$, $\theta$, $\phi$) with the inner and outer boundaries at 1 and 80 AU, respectively. While other groups have also developed 3-D multi-fluid models to investigate the evolution of large-scale solar wind [e.g., \cite{Usmanov2014ApJ,vanderHolst2010ApJ,Sokolov2013ApJ}], we take a different approach in assembling long-term time-varying boundary conditions.

We use OMNI 1-day averaged plasma and $|\textbf{B}|$ data from 01 January 1995 to 31 December 2015 to partially fill the inner boundary surface at 1-day intervals. For each boundary frame, we fill the equatorial band with OMNI data from up to 13 days ahead of and behind the particular day under the assumption of a co-rotating solar wind. In other words, we place the current day's data at the central meridian and then fill the region west (or east) of the central meridian with the past (or future) 1--13 days' data at 13.2\textdegree\ intervals. On the far side, the data from 13 days in the past and future are averaged to prevent an unrealistic jump across the middle. While the use of OMNI plasma data is relatively straightforward, magnetic field components are rather difficult because individual averaged components are often unreasonably small due to rapid fluctuations between positive and negative values during the averaging period, especially when Earth moves across the heliospheric current sheet (HCS). Therefore, we compute magnetic field components from OMNI $|\textbf{B}|$ data in the form of a Parker spiral field with latitudinal dependence. Furthermore, we assume a unipolar magnetic field configuration in order to eliminate numerical dissipation across the HCS. In principle, the correct polarity can be retrieved afterwards because we track the HCS surface using a level set method \citep{Borovikov2011ApJL}.

Outside the equatorial band filled by OMNI data, we construct idealized polar coronal holes (PCHs) varying in size as a function of time [e.g., \cite{HarveyRecely2002SoPh}]. For example, the PCH (or OMNI band) boundaries, which initially reached down to $\pm$30\textdegree\ (or $\pm$10\textdegree) in 1995 around solar minimum, gradually retreated toward the poles to $\pm$80\textdegree\ (or $\pm$60\textdegree) in 2001 around solar maximum. They then gradually descended to $\pm$40\textdegree\ (or $\pm$20\textdegree) in 2008 at the next solar minimum before climbing back to $\pm$80\textdegree\ (or $\pm$60\textdegree) in 2014 at the recent solar maximum. In the 20\textdegree-wide buffer region bounded by the PCH and OMNI band edges, we linearly interpolate between the PCH and OMNI values to avoid unreasonable jumps and to better match Ulysses data that changed rather gradually at those latitudes.

The model PCHs are centered at the poles and are symmetric about the equatorial plane, but in reality, they are usually asymmetric and tilted. Furthermore, PCHs sometimes have large extensions to the equatorial region as they did in 2003 [see \cite{Elliott2012JGR}]. At the inner boundary, the solar wind velocity changes from 800 km s$^{-1}$ at the center (poles) to $\sim$700 km s$^{-1}$ at the edges of PCHs. Empirical correlations from Ulysses data are used to estimate the PCH density and temperature as a function of solar wind radial velocity \citep{Ebert2009JGR,Pogorelov2013ApJ} using the following formulae:
\begin{itemize}
\item
SC 22 (Fast wind: $v_R$ $>$ 500 km s$^{-1}$)
\begin{equation}
\begin{split}
T=&1.2\times 10^5 R^{-1.02} [2.58 + 0.00234\times\\
  &(v_R R^{-0.06} - 706)],\\
\end{split}
\end{equation}
\begin{equation}
\begin{split}
n=&R^{-1.93} [3.07 - 0.0116\times\\
  &(v_R R^{-0.06} - 706)],
\end{split}
\end{equation}
and
\item
SC 23 (Fast wind: $v_R$ $>$ 450 km s$^{-1}$)
\begin{equation}
\begin{split}
T=&1.2\times 10^5 R^{-0.95} [2.21 + 0.000233\times\\
  &(v_R R^{-0.06} - 668)],\\
\end{split}
\end{equation}
\begin{equation}
\begin{split}
n=&1.2\times R^{-1.93} [2.27 - 0.0085\times\\
  &(v_R R^{-0.06} - 668)],
\end{split}
\end{equation}
\end{itemize}
where $v_R$, $R$, $T$, and $n$ are the solar wind radial velocity (km s$^{-1}$), heliocentric distance (AU), proton temperature (K), and number density (cm$^{-3}$), respectively. We note that the original formulae were slightly modified to better match our solutions at PCH latitudes to Ulysses data. Finally, we compute magnetic field components in the form of a Parker spiral field from running solar rotation averages of OMNI $|\textbf{B}|$ data, based on the assumption that the radial field component is invariant and independent of heliographic latitude as suggested by Ulysses data \citep{SmithBalogh1995GRL,Smith2001GRL}.

At the outer boundary at 80 AU, the interstellar hydrogen atom density is 0.09 cm$^{-3}$ in the model, which matches the estimated value of 0.09 $\pm$ 0.022 cm$^{-3}$ at the nose of the termination shock \citep{Bzowski2009SSRv}. We use the inflow speed, direction, and temperature of interstellar hydrogen suggested by \cite{McComas2015ApJS} based on IBEX observations - i.e., 25.4 km s$^{-1}$, 75.7\textdegree\ ecliptic inflow longitude, -5.1\textdegree\ ecliptic inflow latitude, and 7500 K, respectively.

%% In this section, we use  the \subsection command to set off
%% a subsection.  \footnote is used to insert a footnote to the text.


%% This section also includes several of the displayed math environments
%% mentioned in the Author Guide.

\section{Results}

\begin{figure}[h]
\begin{center}
\noindent\includegraphics[width=0.5\textwidth, angle=0]{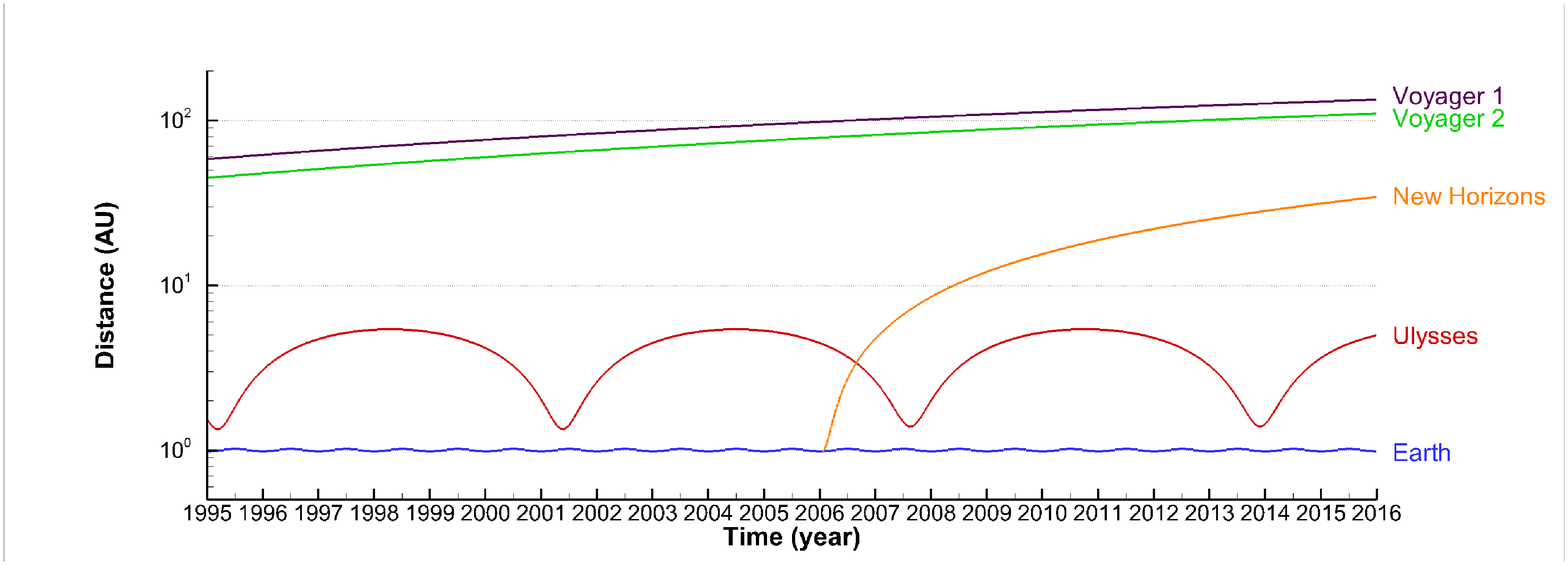}
\noindent\includegraphics[width=0.5\textwidth, angle=0]{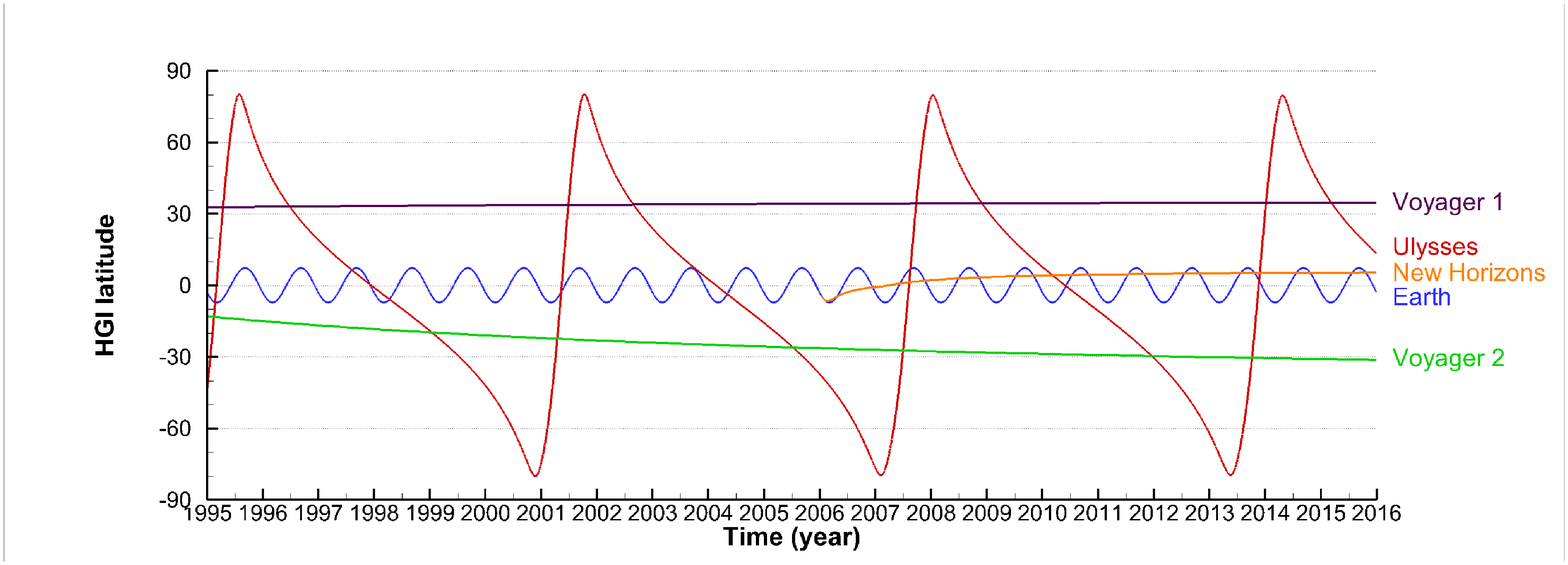}
\noindent\includegraphics[width=0.5\textwidth, angle=0]{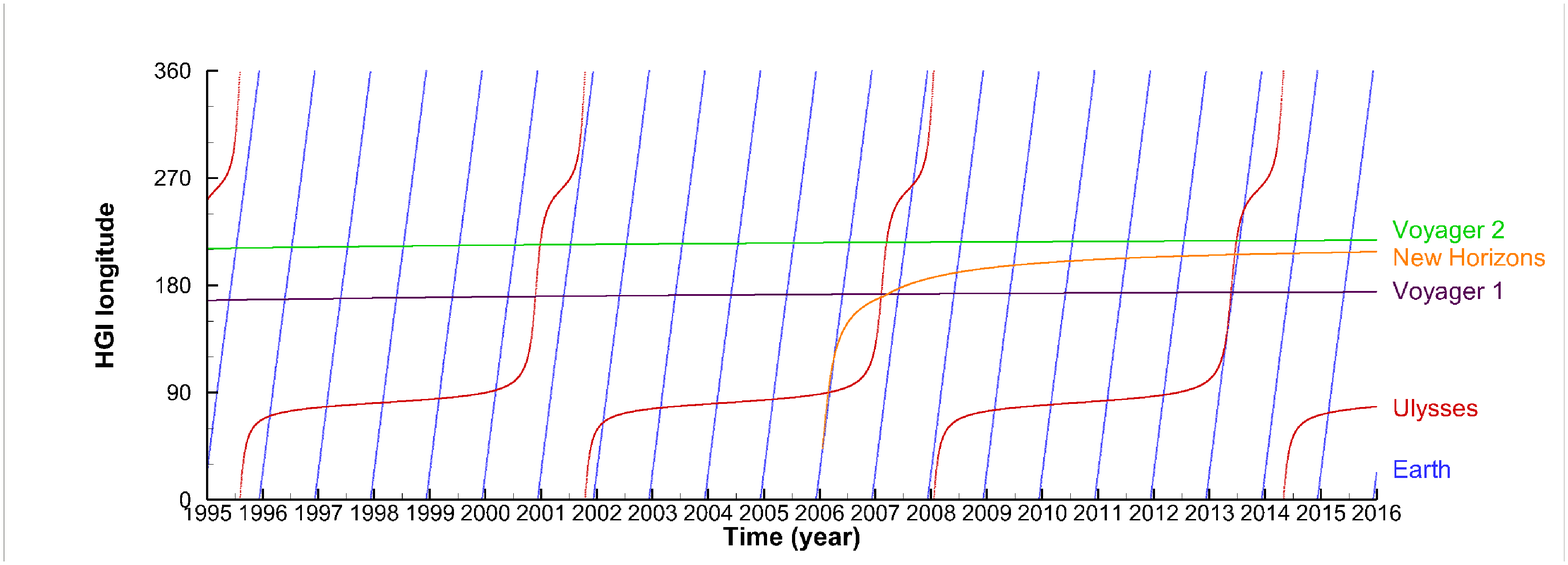}
\end{center}
\caption{The trajectories of Earth, Ulysses, Voyager, and NH from 1995 to 2015 are shown in the order of heliocentric distance (top), latitude (middle), and longitude (bottom) in the HGI coordinate system.}
\label{trajectory}
\end{figure}

The trajectories of Ulysses, Voyager, and NH lie within the simulation domain allowing us to directly compare our solutions with spacecraft data. Figure \ref{trajectory} shows the heliocentric distance, latitude, and longitude in the heliographic inertial (HGI) coordinate system as a function of time. The Ulysses mission ended on 30 June 2009 while Voyager 1 (V1) and 2 (V2) moved beyond the outer boundary around 2001.0 and 2006.5, respectively. Thus, we provide comparisons only up to those dates. Moreover, we compare proton temperatures only at Ulysses in the inner heliosphere, but not at Voyager and NH in this paper because the model values reflect the temperature of an isotropic mixture of solar wind and PUIs rather than that of the solar wind by itself. It is also important to note that NH's latitude changes asymptotically from -6.6\textdegree\ shortly after launch to +5.3\textdegree\ at Pluto, remaining very close to the ecliptic plane. 

\subsection{Comparison with Ulysses Data}

\begin{figure}[h]
\begin{center}
\noindent\includegraphics[width=0.5\textwidth, angle=0]{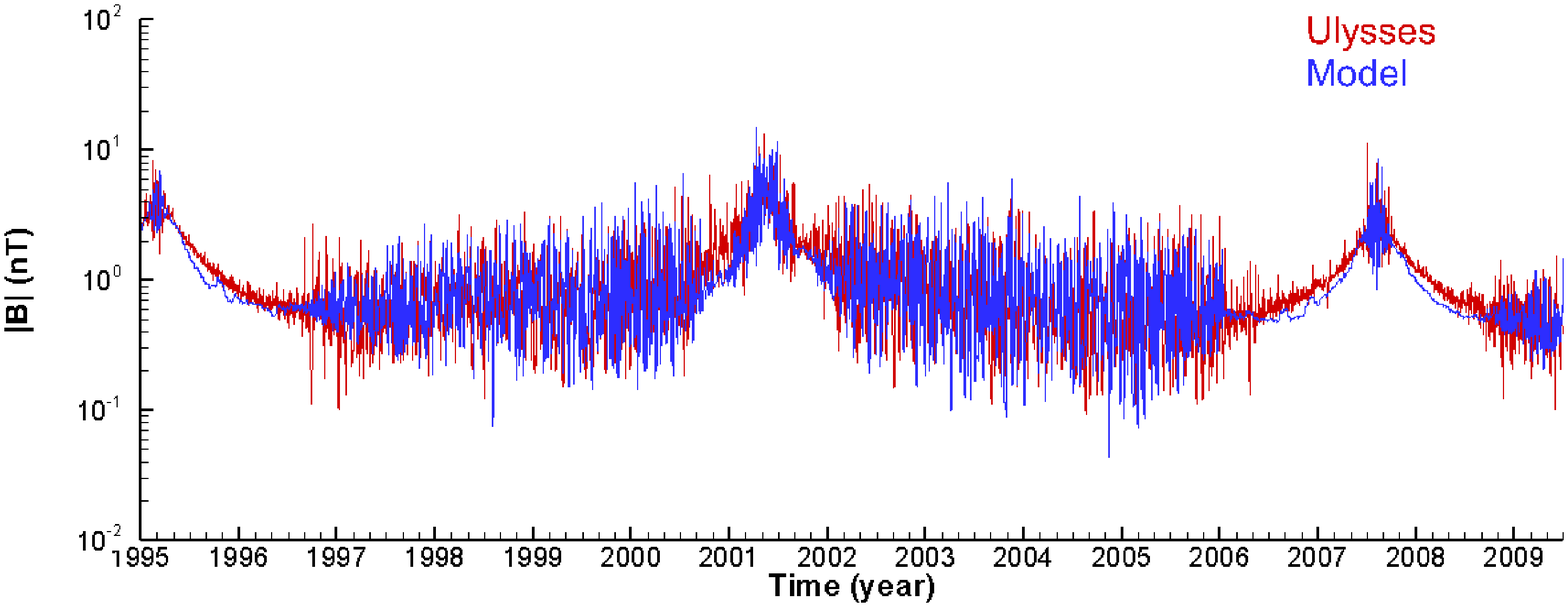}
\noindent\includegraphics[width=0.5\textwidth, angle=0]{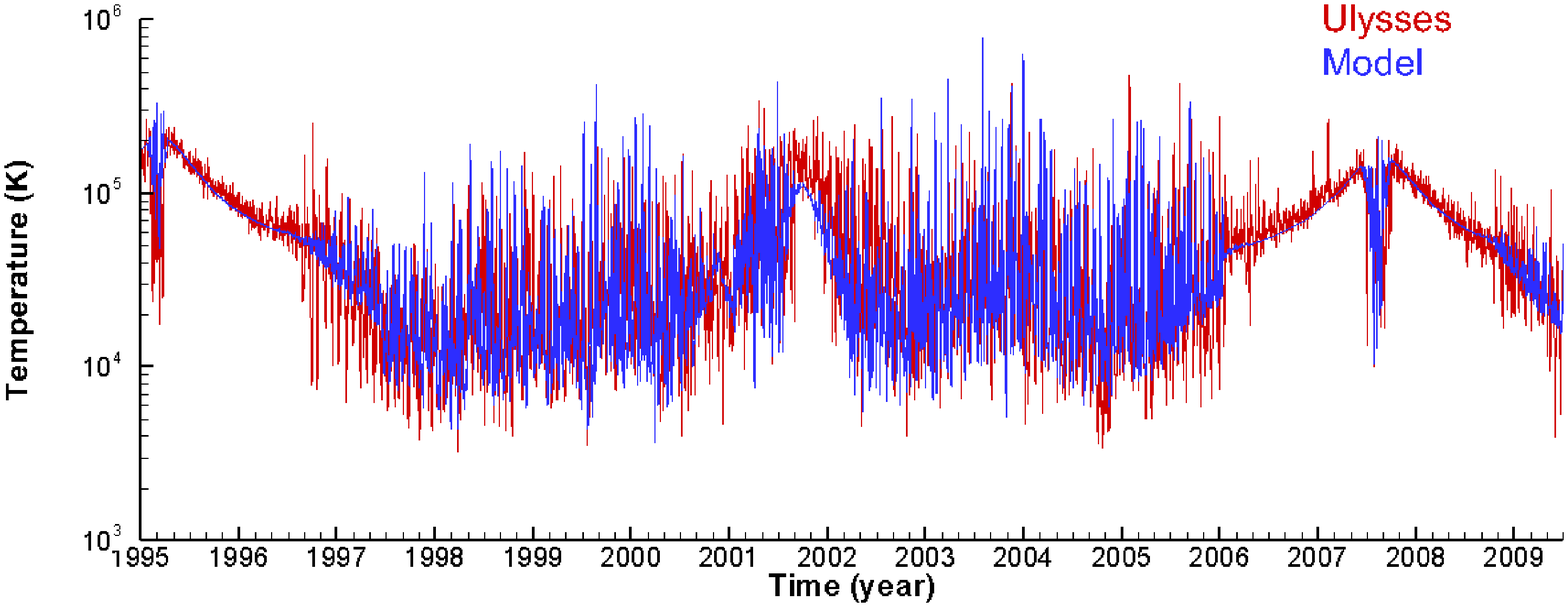}
\noindent\includegraphics[width=0.5\textwidth, angle=0]{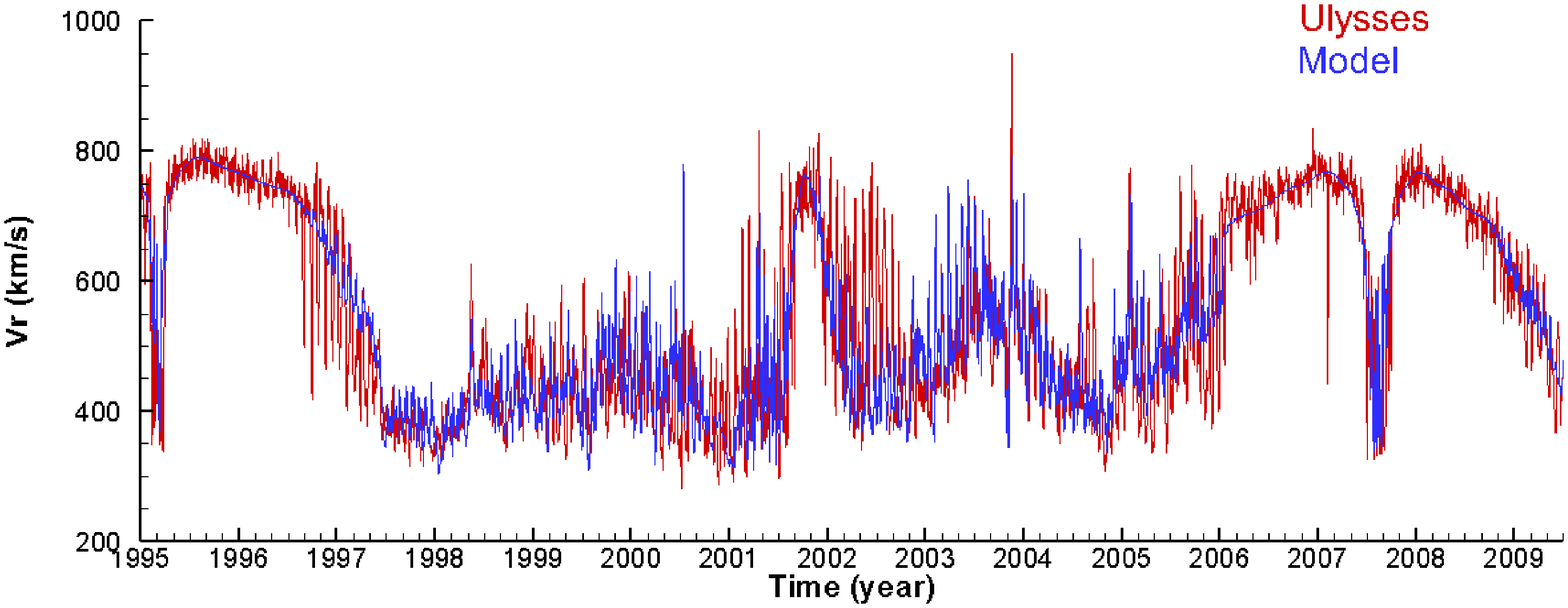}
\noindent\includegraphics[width=0.5\textwidth, angle=0]{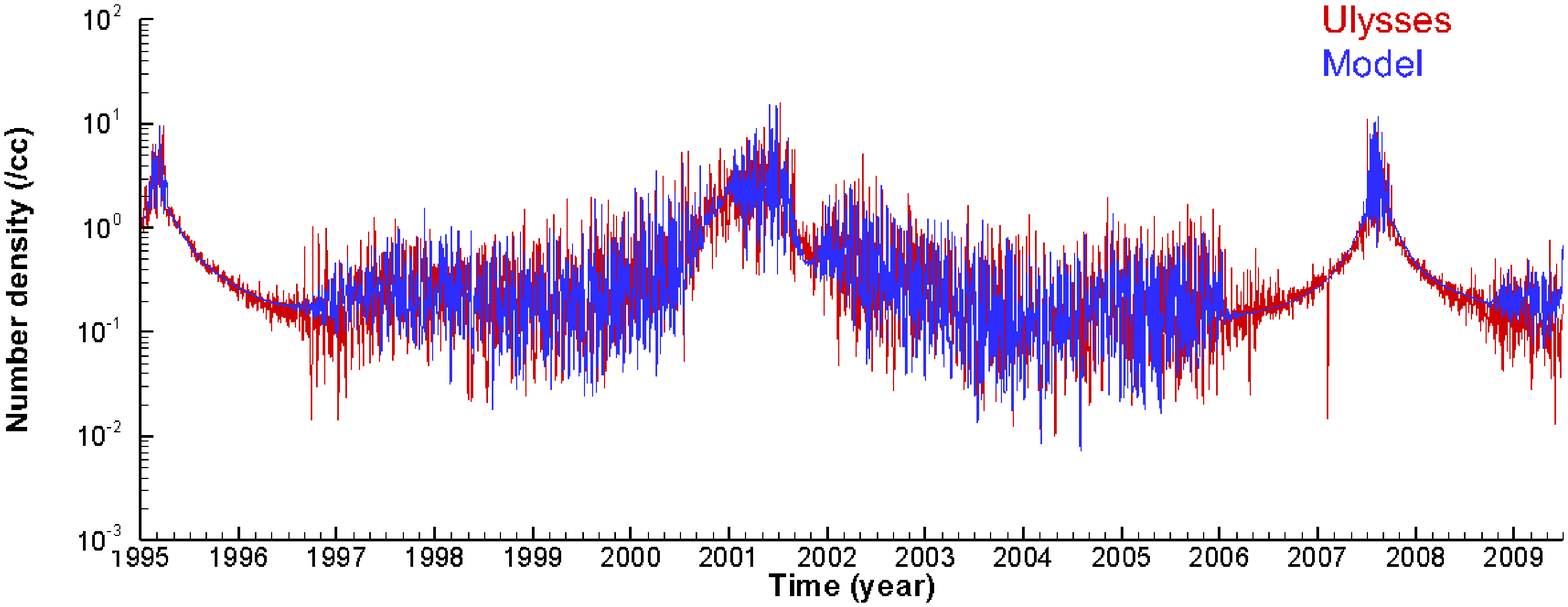}
\end{center}
\caption{Model $|\textbf{B}|$, proton temperature, radial velocity, and number density between 1995.0 and 2009.5 are compared with the daily averaged Ulysses data.}
\label{Ulysses}
\end{figure}

Figure \ref{Ulysses} compares model $|\textbf{B}|$, proton temperature, radial velocity, and density with Ulysses data. For most of the 15-year period, our model matches Ulysses data reasonably well outside the latitudinal extent of PCHs where the model reflects fluctuations in OMNI data. Between 1995.0 and 1995.5 at SC 22 solar minimum, Ulysses made a fast transit across the equatorial plane from -45\textdegree\ to +80\textdegree\ latitude. Fluctuations in $|\textbf{B}|$, temperature, and density are relatively large within $\pm$30\textdegree\ of the equatorial plane where the radial velocity drops below 400 km s$^{-1}$. At PCH latitudes, the simulation results vary much more smoothly than do spacecraft data mainly because the boundary conditions at 1 AU do not contain Alfv\'{e}nic fluctuations and change smoothly with latitude in PCHs. The same pattern is observed during another pole-to-pole fast transit between 2007.0 and 2008.0 near SC 23 solar minimum, although the PCHs were characterized by slightly slower, less dense, cooler solar wind and weaker polar magnetic fields than in the SC 22 solar minimum \citep{McComas2008GRL}.

Around 2000.9 during solar maximum, Ulysses data exhibit wide variability even at high latitudes due to the complex global solar wind structure \citep{McComas2002GRL}. While our model is not expected to reproduce such large fluctuations as seen at Ulysses near the poles owing to the simple geometry of the boundary conditions, it does manage to replicate the long-term variations over the period. A similar pattern is observed around 1997.0 and 2006.0 during which Ulysses traverses mid-latitude regions between PCHs and the equatorial band characterized by the interaction of slow wind and PCH fast wind streams [e.g., \cite{Gosling1996ARAA}]. Occasional coronal mass ejection (CME) and comet tail passages also contributed to the discrepancy between the model and spacecraft data [see \cite{Ebert2009JGR}, \cite{Elliott2012JGR} and references therein]. The model $|\textbf{B}|$ at latitudes above $\pm$60\textdegree--70\textdegree\ is also systematically lower than Ulysses data by up to 30 \% suggesting that we may have underestimated magnetic field values near the poles at the inner boundary, but an accurate polar magnetic field estimation is beyond the scope of our study.

\begin{figure}[H]
\begin{center}
\noindent\includegraphics[width=0.38\textwidth, angle=0]{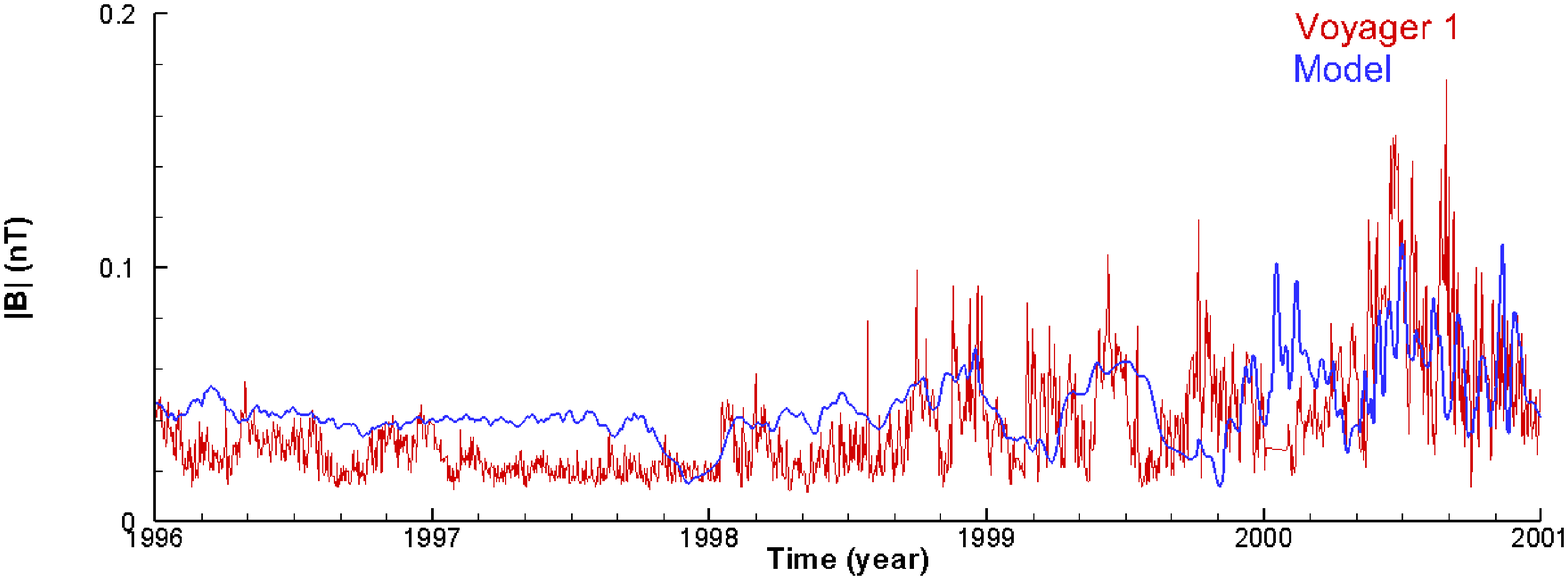}
\noindent\includegraphics[width=0.38\textwidth, angle=0]{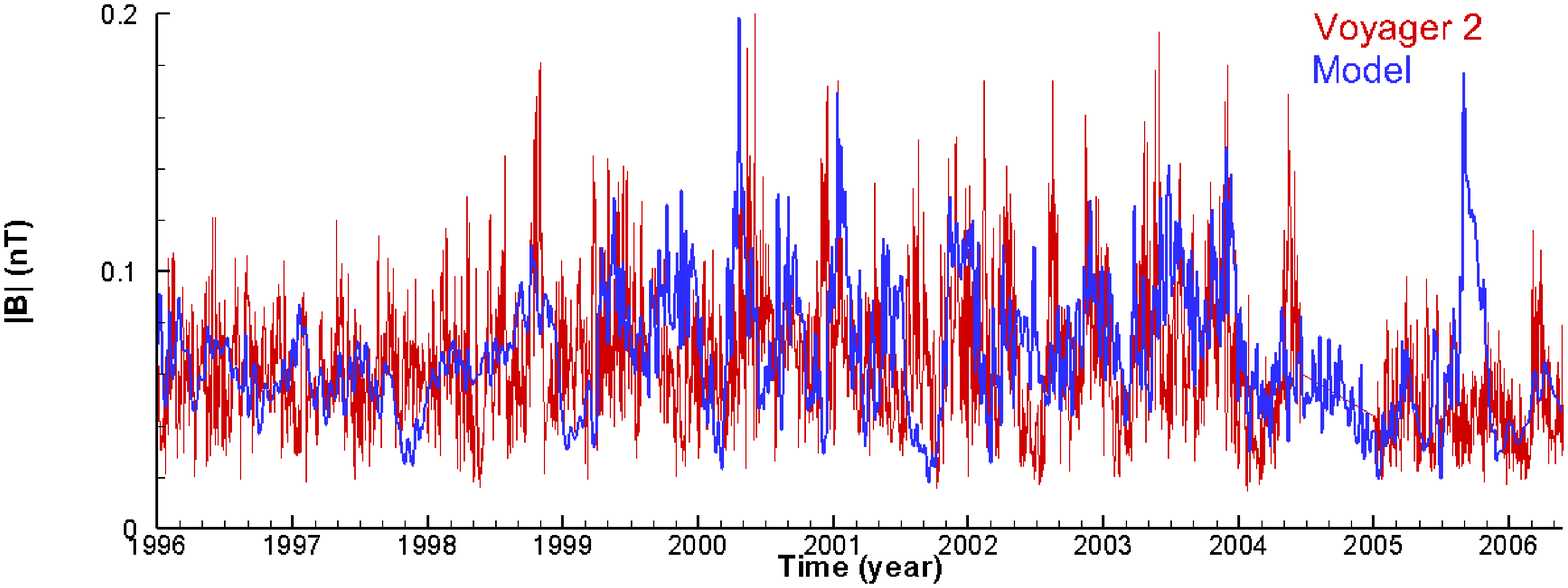}
\noindent\includegraphics[width=0.38\textwidth, angle=0]{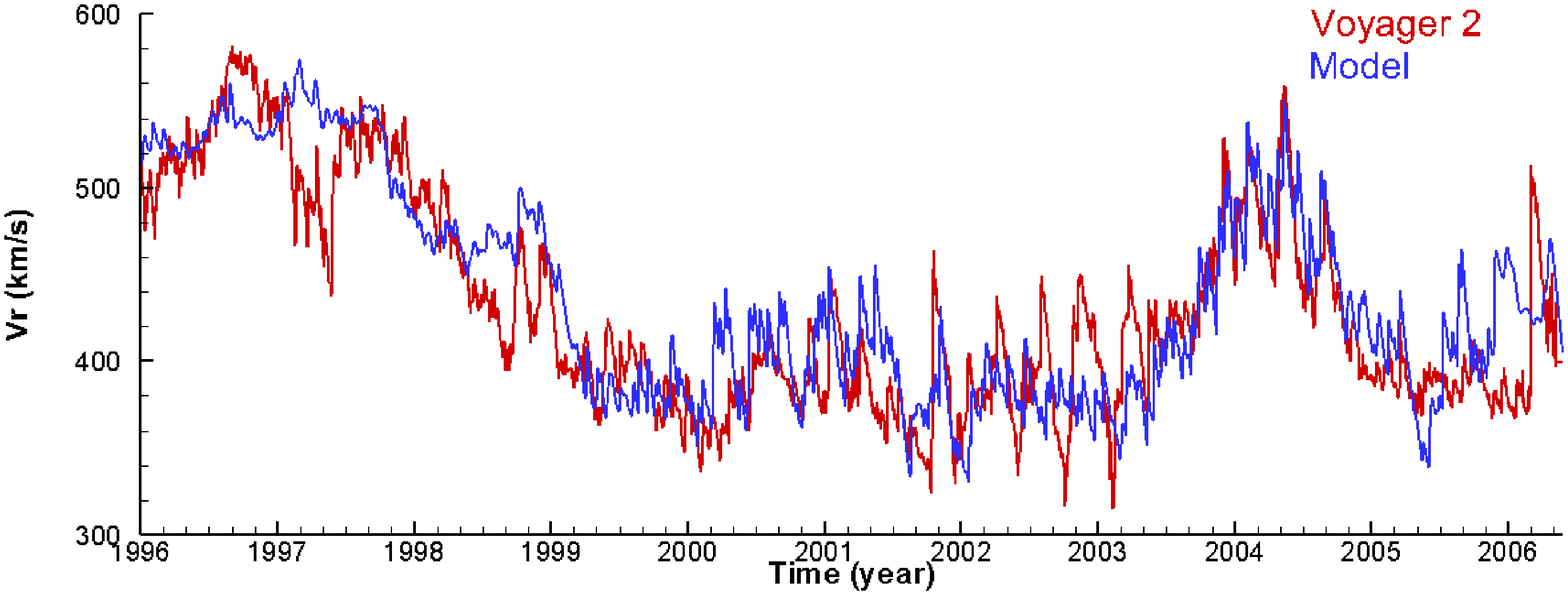}
\noindent\includegraphics[width=0.38\textwidth, angle=0]{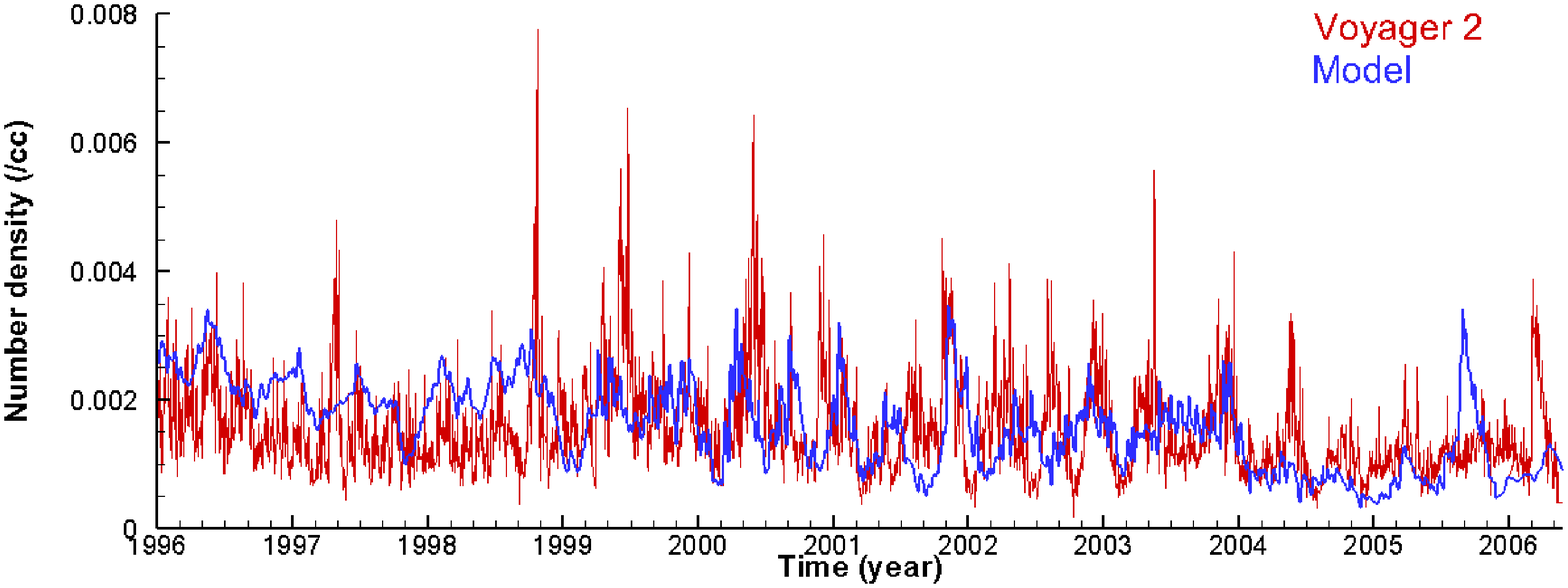}
\noindent\includegraphics[width=0.38\textwidth, angle=0]{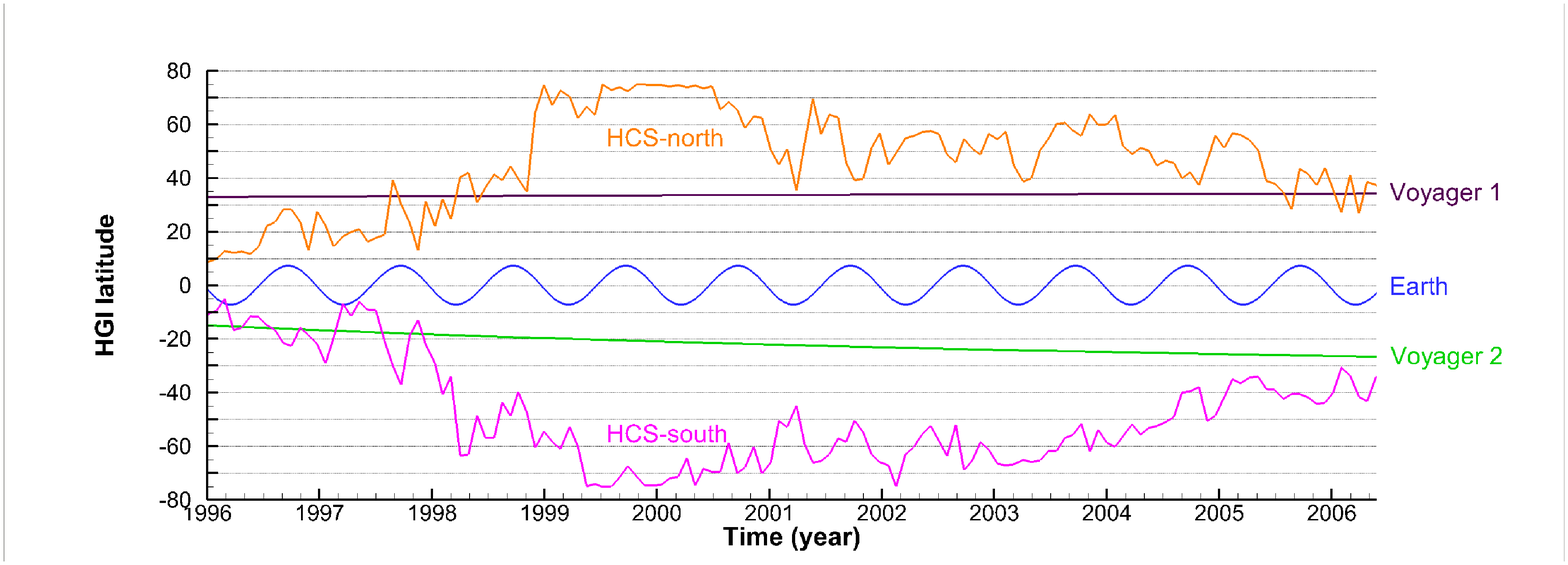}
\end{center}
\caption{Model $|\textbf{B}|$ at V1 and V2, radial velocity and number density at V2 are compared to the daily averaged spacecraft data. In the bottom plot, Carrington rotation averages of the maximum HCS inclination (from WSO) in both hemispheres are shown as a function of time along with HGI latitudes of V1, V2, and Earth.}
\label{Voyager}
\end{figure}

\subsection{Comparison with Voyager Data}

Figure \ref{Voyager} compares the model $|\textbf{B}|$, radial velocity, and proton density with Voyager spacecraft data. Since V1 plasma data are not available, we compare only $|\textbf{B}|$ at V1. Overall, our model reproduces the gradual increase in $|\textbf{B}|$ observed by V1, which is likely due to solar cycle and latitude variations \citep{Burlaga2002JGR}, and the relatively large fluctuations in $|\textbf{B}|$ at V2. From 1996 to 1998, the model $|\textbf{B}|$ at V1 is nearly steady around 0.04 nT while the observed $|\textbf{B}|$ fluctuates between 0.02 and 0.05 nT. Inaccuracies in our estimates of $|\textbf{B}|$ in mid-latitude PCH fast streams may have contributed to a slightly larger model $|\textbf{B}|$ at +33\textdegree\ than observed at V1 during this period. From 1998 to 2001, fluctuations grow steadily as OMNI data begin to exert increasing influence on the model $|\textbf{B}|$ at V1 while the PCH retreats to higher latitudes. For much of the 5-year period, the model $|\textbf{B}|$ values at V1 are within the instrument's measuring uncertainty of 0.02 nT.

On the other hand, V2 moved from -15\textdegree\ to -25\textdegree\ and observed relatively large fluctuations in $|\textbf{B}|$ throughout the 10-year period shown in Figure \ref{Voyager}. The model accurately reproduces the large-scale increase in $|\textbf{B}|$ fluctuations toward solar maximum around 2000 and subsequent decrease toward solar minimum, although it contains an anomalous jump around 2005.65 that was not observed by V2. Accompanied by a prominent density peak, the large width of this structure hints at a possible merged interaction region that somehow did not materialize at V2 as predicted by the model. This is hardly surprising given the latitudinal offset of 20\textdegree--30\textdegree\ between Earth and V2 at that time.

While the model velocity changes relatively smoothly between 1996 and 1999, the observed velocity plummeted by more than 100 km s$^{-1}$ before recovering higher values in 1997. This velocity dip in early 1997 may be explained by the latitudinal extent of the solar wind being affected by the time-varying HCS tilt angle (see Figure \ref{Voyager}) that changed from -10\textdegree\ to -30\textdegree\ during 1996 and back to -10 degrees around 1997.2, according to estimates by the Wilcox Solar Observatory (WSO). V2 moved from -15\textdegree\ to -17\textdegree\ latitude during that time, so it may have encountered slow wind streams around the HCS when the tilt angle remained within $\pm$5\textdegree of its latitude between 1996.2 and 1997.0. In fact, \cite{RichardsonWang1998GRL} have shown a strong anti-correlation between HCS tilt angle and solar wind speed observed by V2 from 1995 to 1998. We also identify sizable discrepancies between the model velocity and V2 data during 1998--1999 and 2005--2006 where model values are consistently higher by up to 80 and 100 km s$^{-1}$, respectively. We note that the model velocities at those times were partly affected by PCH fast streams because V2 was in the transition region between PCH and OMNI-driven equatorial band, but V2 observations suggest little influence, if at all, by PCH streams [e.g., \cite{BurlagaNess2000JGR}]. Limited by simple geometry of the boundary values, we do not expect our model to reproduce HCS-related structures or stream interactions at the interface between PCH fast wind and equatorial solar wind streams with high precision.

\subsection{Comparison with NH SWAP Data}

\begin{figure}[H]
\begin{center}
\noindent\includegraphics[width=0.41\textwidth, angle=0]{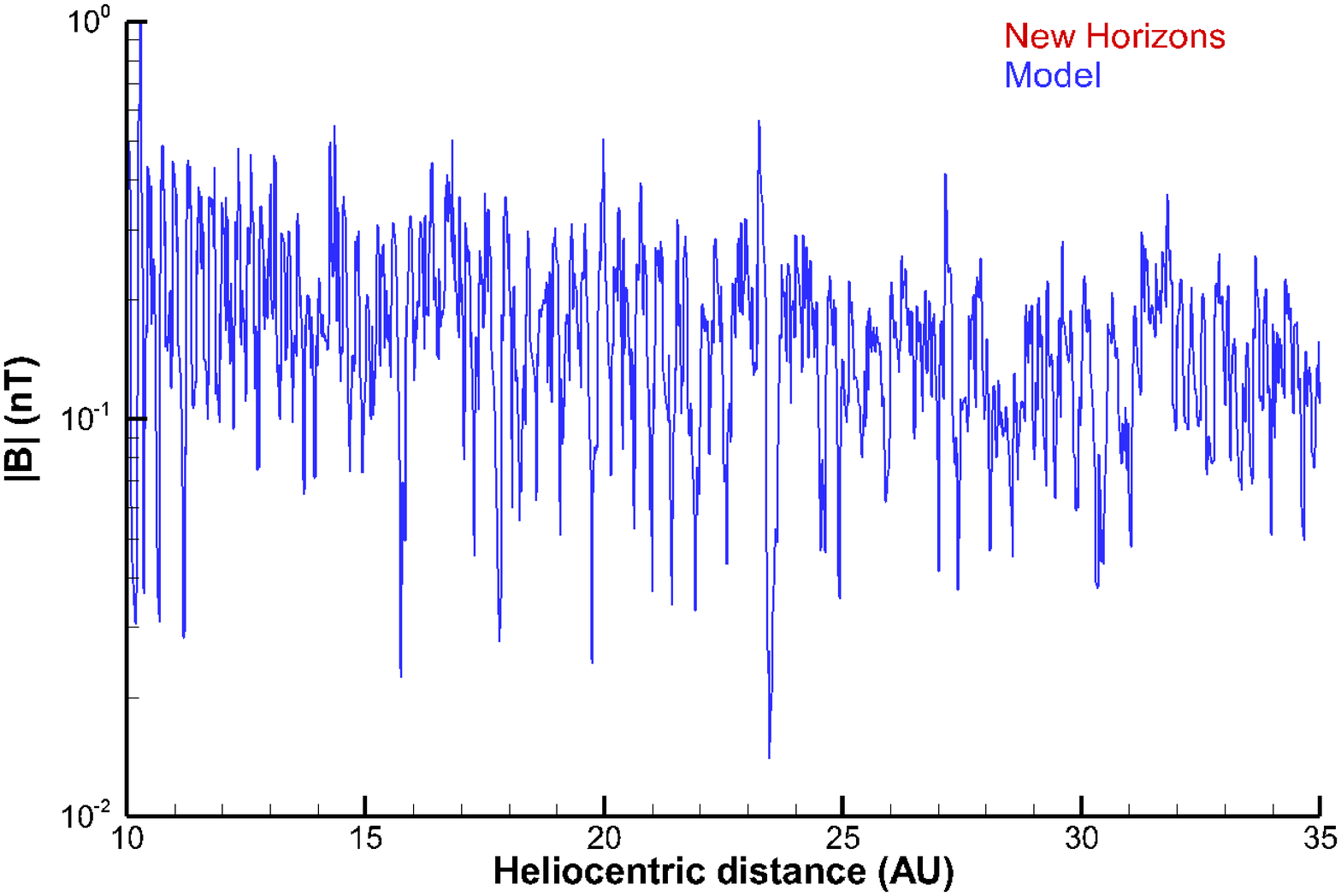}
\noindent\includegraphics[width=0.41\textwidth, angle=0]{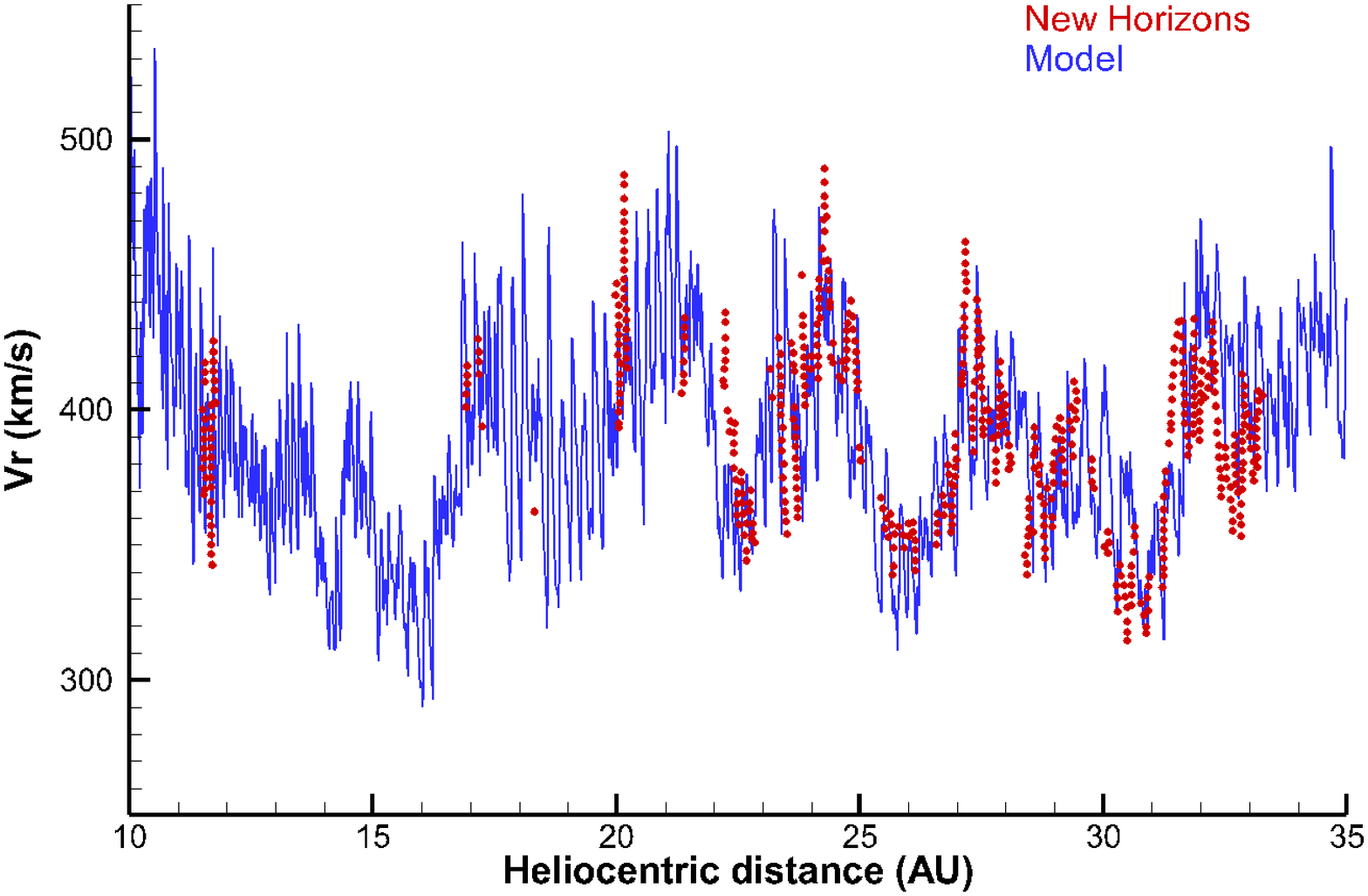}
\noindent\includegraphics[width=0.41\textwidth, angle=0]{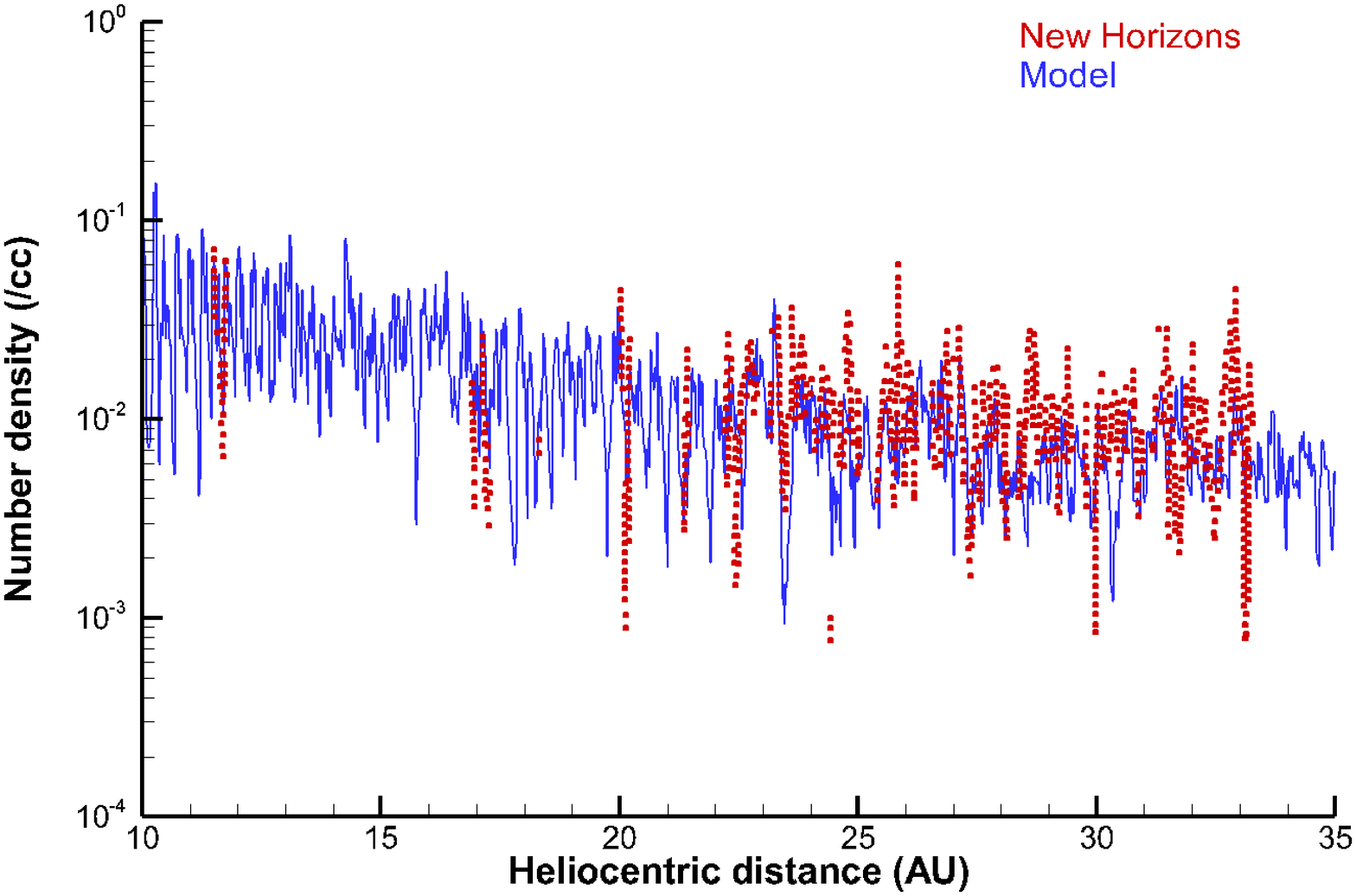}
\end{center}
\caption{Model $|\textbf{B}|$, radial velocity, and number density are compared to the daily averaged NH-SWAP data. Adapted from \cite{Elliott2016ApJS} with permission of the AAS.}
\label{NH}
\end{figure}

In Figure \ref{NH}, we show the model $|\textbf{B}|$, radial velocity, and proton density along NH's trajectory between 10 and 35 AU from the Sun. While we cannot compare with in situ $|\textbf{B}|$ measurements because NH is not equipped with a magnetometer, we compare the radial velocity and proton density with recently published solar wind observations from SWAP \citep{Elliott2016ApJS}. The SWAP design is described by \cite{McComas2008SSRv}.

The model $|\textbf{B}|$ fluctuates mostly between 0.05 and 0.30 nT at these distances, which is consistent with estimates from \cite{Bagenal2015JGR} based on V2 measurements \citep{Richardson2003GRL}. The radial velocity varies steeply between 300 and 600 km s$^{-1}$ at 1 AU, but the fluctuations at both small and large scales are considerably reduced with distance, in agreement with SWAP data \citep{Elliott2016ApJS}. While stream interaction between parcels of wind with different speed is responsible for gradually removing small and mid-size structures \citep{Gazis1995GRL,Richardson1996AIP,Richardson2002JGR,Elliott2016ApJS}, large scale fluctuations are affected by changes related to solar cycle. For example, a number of persistent, low-latitude coronal holes during the SC 23 minimum were associated with recurring fast wind streams of $\sim$600 km s$^{-1}$ at 1 AU \citep{Abramenko2010ApJ}, but such fast streams appear less frequently in OMNI data after 2009 in SC 24. Lastly, the model proton density steadily decreases out to 20 AU and then fluctuates around 0.01 cm$^{-3}$ between 20 and 33 AU, in agreement with SWAP data. However, the fluctuations in SWAP density are markedly larger than the model values at certain distances between 20 and 33 AU (i.e., 20.1, 25.8, 28.5, 31.7, and 33.0 AU). We point out that these discrepancies occur when the longitudinal separation between Earth and NH is relatively large, during a period of increased solar activity between 2011 and 2015. It may be possible to improve the accuracy of our model for this period by incorporating STEREO (Solar Terrestrial Relations Observatory) data in the boundary conditions to relax the assumption of co-rotating solar wind at 1 AU.

In Table 1, we compare the model with SWAP data at Pluto. The model solar wind speed, proton density, flux, and ram pressure are 448 km s$^{-1}$, 0.012 cm$^{-3}$, 5.4 km s$^{-1}$ cm$^{-3}$, and 3.6 pPa, whereas the measured values are 403 km s$^{-1}$, 0.025 cm$^{-3}$, 10 km s$^{-1}$ cm$^{-3}$, and 6.0 pPa, respectively \citep{Bagenal2016Science,McComas2016JGR}. The model predicts an interplanetary magnetic field strength of 0.2 nT which is somewhat lower than 0.3 nT assumed by \cite{Bagenal2016Science}, but we note that the model value falls within the range of estimated values 0.08--0.30 nT. The difference in model and observed speed is relatively small compared to the difference between proton densities. \cite{Bagenal2016Science} report that a strong interplanetary shock arrived at NH on 09 July 2015, which may explain the unusually large solar wind density measured by SWAP. On the other hand, the model shows a modest jump in $|\textbf{B}|$, velocity, and density on 10 July 2015 when a complicated structure associated with a CME that caused the geomagnetic storm of 17--18 March 2015 (the St. Patrick's Day storm) \citep{Astafyeva2015JGR} reaches NH. On 17 March 2015, NH was 108\textdegree\ ahead of Earth longitudinally. Observations by the Large Angle and Spectrometric Coronagraph instrument aboard the Solar and Heliospheric Observatory satellite indicate that the center of the halo CME was initially directed at an angle between the longitude coordinates of Earth and NH, but it may have been further deflected toward NH by a trailing fast wind stream. It is possible that NH encountered a very different part of the CME-related structure with greater enhancement in density as it passed by Pluto.

\begin{table}[H]
\begin{center}
\caption{Comparison with Solar Wind Observations at Pluto \citep{Bagenal2016Science,McComas2016JGR}.\label{model-obs-comp}}
\begin{tabular}{crrrrrrrrrrr}
\tableline\tableline
Parameter & Model & Observation\\
\tableline
Solar wind speed (km s$^{-1}$) &448 &403\\
Proton density (cm$^{-3}$) &0.012 &0.025\\
Proton flux (km s$^{-1}$ cm$^{-3}$) &5.4 &10\\
Proton ram pressure (pPa) &3.6 &6.0\\
Magnetic field strength (nT) &0.2 &0.3\tablenotemark{a}\\
\tableline
\end{tabular}
\tablenotetext{a}{The maximum value derived from V2 observations.}
\end{center}
\end{table}

\section{Summary and Discussion}

Using daily averaged solar wind parameters from OMNI data in combination with idealized PCHs to fit Ulysses data at high latitudes in the inner heliosphere, we modeled a 3-D time-varying heliosphere out to 80 AU. Our model accurately reproduced the long-term variations observed by Ulysses, Voyager, and NH. However, comparisons with Ulysses and Voyager data show systematic discrepancies at the interface between OMNI-driven equatorial region and PCHs because the PCH boundaries in reality are not perfect arcs as assumed by the model, but have extensions. We may improve our model at Ulysses and Voyager by introducing at the inner boundary a dipole magnetic field with a tilted solar wind structure consistent with the observed time-varying HCS tilt angle. This would allow us to extend the simulation into the heliosheath and beyond to investigate shock propagation across the heliopause in a future study.

We did not compare proton temperatures at Voyager and NH in this paper because the model values reflect the temperature of an isotropic mixture of solar wind and PUIs rather than that of the solar wind by itself. In a follow-up study, we will use a more sophisticated version of MS-FLUKSS consisting of a turbulence model and a three-fluid model treating PUIs as a separate fluid from the solar wind [e.g., \cite{Kryukov2012AIP} and \cite{Usmanov2014ApJ}]. In addition to PUI and turbulence parameters, we expect the model to reproduce, with reasonable accuracy, the solar wind temperature in the outer heliosphere where PUI-driven turbulence contributes to significant solar wind heating. We will continue our modeling effort to provide comparison with the exciting measurements of NH as the mission extends beyond Pluto.


%% Included in this acknowledgments section are examples of the
%% AASTeX hypertext markup commands. Use \url without the optional [HREF]
%% argument when you want to print the url directly in the text. Otherwise,
%% use either \url or \anchor, with the HREF as the first argument and the
%% text to be printed in the second.

\acknowledgments

This research was inspired by the New Horizons Flyby Modeling Challenge hosted by CCMC/NASA. The authors acknowledge use of the SPDF COHOWeb database and WSO data (\url{wso.stanford.edu}). This work is supported by the NSF PRAC award OCI-1144120 and related computer resources from the Blue Waters sustained-petascale computing project. Supercomputer time allocations were also provided on SGI Pleiades by NASA High-End Computing Program award SMD-15-5860 and on Stampede by NSF XSEDE project MCA07S033. T. K. Kim and N. V. Pogorelov acknowledge support from the NSF SHINE project AGS-1358386, and NASA projects NNX12AB30G and NNX14AF41G.

%% Tables may also be prepared as separate files. See the accompanying
%% sample file table.tex for an example of an external table file.
%% To include an external file in your main document, use the \input
%% command. Uncomment the line below to include table.tex in this
%% sample file. (Note that you will need to comment out the \documentclass,
%% \begin{document}, and \end{document} commands from table.tex if you want
%% to include it in this document.)


%% The following command ends your manuscript. LaTeX will ignore any text

\end{document}